\documentclass[aps,twocolumn,prd,reprint,superscriptaddress,nofootinbib,floatfix]{revtex4-2}
\RequirePackage[l2tabu, orthodox]{nag}
\usepackage{lipsum}

\usepackage[utf8x]{inputenc}
\usepackage{microtype}

\usepackage{amsmath}
\usepackage{amstext, amssymb, amsthm, amsfonts, slashed}
\usepackage{mathtools}
\usepackage[version=4]{mhchem}
\usepackage{graphicx}
\usepackage{booktabs}
\usepackage{dcolumn}
\usepackage{bm}
\usepackage{dsfont}
\usepackage[usenames,dvipsnames]{xcolor}
\usepackage[normalem]{ulem}

\usepackage{url}
\usepackage[colorlinks=true,urlcolor=blue,linkcolor=blue,citecolor=blue,hypertexnames]{hyperref}
\usepackage[nameinlink]{cleveref}
\usepackage{braket}
\usepackage{simplewick}
\usepackage{float}

\hypersetup{
	colorlinks,
	linkcolor={blue!75!black},
	citecolor={blue!75!black},
	urlcolor={blue!75!black}
}

\crefname{section}{Sec.\!}{Secs.\!}
\crefname{figure}{Fig.\!}{Figs.\!}
\crefname{equation}{}{}
\crefname{table}{Tab.\!}{Tabs.\!}
\crefname{appendix}{App.\!}{Apps.\!}

\def\s0#1#2{\mbox{\small{$ \frac{#1}{#2} $}}}
\def\0#1#2{\frac{#1}{#2}}
\newcommand*\dif{\mathop{}\!\mathrm{d}}
\newcommand{\Tr}{\mathrm{Tr}}

\newcommand{\imag}{\text{i}}
\DeclareMathOperator{\arccosh}{arcosh}

\graphicspath{{figures/}}

\def\CG{{\mathcal G}}

\def\CR{{\mathcal R}}

\begin{document}

\title{Lorentzian quantum gravity and the graviton spectral function}

\author{Jannik~Fehre}
\affiliation{Institut für Theoretische Physik, Universität Heidelberg, Philosophenweg 16, 69120 Heidelberg, Germany}
\author{Daniel~F.~Litim}
\affiliation{Department  of  Physics  and  Astronomy,  University  of  Sussex,  Brighton,  BN1  9QH,  U.K.}
\author{Jan~M.~Pawlowski}
\affiliation{Institut für Theoretische Physik, Universität Heidelberg, Philosophenweg 16, 69120 Heidelberg, Germany}
\affiliation{ExtreMe Matter Institute EMMI, GSI Helmholtzzentrum f\"ur Schwerionenforschung mbH, Planckstr.\ 1, 64291 Darmstadt, Germany}
\author{Manuel~Reichert}
\affiliation{Department  of  Physics  and  Astronomy,  University  of  Sussex,  Brighton,  BN1  9QH,  U.K.}

\begin{abstract}
 We present the first direct and non-perturbative computation of the graviton spectral function in quantum gravity. This is achieved with the help of a novel Lorentzian renormalisation group approach, combined with a spectral representation of correlation functions. We find a positive graviton spectral function, showing a massless one-graviton peak and a multi-graviton continuum with an asymptotically safe scaling for large spectral values. We also study the impact of a cosmological constant. Further steps to investigate scattering processes and unitarity in asymptotically safe quantum gravity are indicated.
 \end{abstract}

\maketitle

\textit{Introduction.---} 
The quest for a consistent quantum theory of gravity continues to offer challenges~\cite{Ashtekar:2014kba}. An important contender is asymptotically safe gravity \cite{Weinberg:1980gg}, where the metric field remains the fundamental carrier of the gravitational force. In this purely quantum field theoretical setup the trans-Planckian ultraviolet regime of quantum gravity is governed by an \textit{interacting} fixed point, and gravity is ruled by the same principles as the Standard Model of particle physics. 

The field of asymptotically safe gravity has seen substantial progress in the past decades, mostly using Euclidean functional renormalisation \cite{Reuter:1996cp}, for reviews see \cite{Litim:2011cp, Bonanno:2017pkg, Eichhorn:2018yfc, Pereira:2019dbn, Reuter:2019byg, Reichert:2020mja, Platania:2020lqb, Bonanno:2020bil, Dupuis:2020fhh, Pawlowski:2020qer}. Nevertheless, the question of unitarity is far from being settled \cite{Bonanno:2020bil, Donoghue:2019clr}, as many results are obtained within Euclidean signature. Naturally, the Wick rotation -- already a subtle issue on flat Minkowski spacetimes -- is further complicated by the dynamical metric. Still, first steps towards computations with Lorentzian signature have been reported \cite{Manrique:2011jc, Rechenberger:2012dt, Demmel:2015zfa, Biemans:2016rvp, Houthoff:2017oam, Wetterich:2017ixo, Knorr:2018fdu, Baldazzi:2018mtl, Nagy:2019jef, Eichhorn:2019ybe, Bonanno:2021squ}, also for other quantum gravity approaches~\cite{Ambjorn:1998xu, Ambjorn:2000dv, Ambjorn:2001cv, Engle:2007uq, Freidel:2007py, Feldbrugge:2017kzv, Asante:2021zzh}. 

In this work, we put forward the first \textit{bona fide} Lorentzian renormalisation group study of asymptotically safe gravity. The key idea is the use of spectral representations for correlation functions, together with an expansion about flat Minkowski spacetime \cite{Pawlowski:2020qer}. In particular, propagators obey the K\"all\'en-Lehmann (KL) representation \cite{Kallen:1952zz, Lehmann:1954xi}. This allows us to find the gravitational fixed point in Lorentzian signature alongside the graviton spectral function. Most notably, the existence of the latter offers access to the graviton propagator for general complex momenta,  including timelike momenta relevant for graviton-mediated scattering processes. \smallskip

\textit{Lorentzian quantum gravity and spectral functions.---} 
We consider Lorentzian quantum gravity based on the classical Einstein-Hilbert action
\begin{align}
	\label{eq:EH-Action}
	S_\text{EH}[g_{\mu\nu}] &= \frac{1}{16 \pi G_\text{N}} \int\! \mathop{\mathrm{d}^{4} x} |\det g_{\mu\nu}|^{\frac12} \,\Bigl(\CR -  2\Lambda \Bigr), 
\end{align}
with Newton's constant $G_\text{N}$, cosmological constant $\Lambda$, and Ricci scalar $\CR[g_{\mu\nu}]$, augmented with a gauge-fixing and ghost action. We use a flat Minkowskian background $\eta=\text{diag}(1,-\bm{1})$ and split the metric field $g_{\mu\nu}= \eta_{\mu\nu}+\sqrt{16 \pi G_\text{N}} \,h_{\mu\nu}$ linearly into background and fluctuation $h_{\mu\nu}$. The main object of interest in the present work is the spectral function of the transverse-traceless (TT) graviton mode with the scalar coefficient $\CG_{hh}$, for which we assume the existence of a KL representation. It relates the spectral function to the propagator via
\begin{align} \label{eq:KS-L}
	\CG_{hh}(p_0,|\vec{p}\,|) = \int_{0}^\infty \frac{\mathrm d\lambda}{\pi}\frac{\lambda\,\rho_h(\lambda,|\vec{p}\,|)}{\lambda^2+p_0^2} \,,
\end{align}
with the temporal and spatial momentum $p_0$ and $\vec{p}$ respectively, the spectral values $\lambda$, and the graviton spectral function
\begin{align}
	\label{eq:rho-G}
	\rho_h(\lambda,|\vec{p}|) = \lim_{\varepsilon\to 0} 2\, \text{Im}\, \CG_{hh}(p_0=-\imag (\lambda + \imag\,\varepsilon),|\vec{p}\,|)\,.
\end{align}
The spectral function acts as a linear response function of the two-point correlator, encoding the energy spectrum of the theory. For asymptotic states, it can be understood as a probability density for the transition to an excited state with energy $\lambda$. The existence of a spectral representation cannot be taken for granted but if it exists, it tightly constrains the analytic structure of the propagator and the asymptotes of the spectral function, see~\cite{Cyrol:2018xeq, Horak:2021pfr} for a discussion in Yang-Mills theories.

It is convenient to  parametrise the spectral function at $\vec p = 0$ through a single-graviton delta-peak with mass $m_h$ and a multi-graviton continuum $f_h$ starting out at the threshold $\lambda=2 m_h$,
\begin{align}\label{eq:rhoh-para}
	\rho_h(\lambda)= \frac{1}{Z_h}\!\Big[ 2 \pi \delta(\lambda^2-m^2_h) +\theta(\lambda^2- 4 m_h^2)  f_h(\lambda)\Big].
\end{align}
Classically, the spectral function is given by a single graviton peak with $m_h=0$ and a trivial wave-function renormalisation, $Z_h=1$. Quantum fluctuations change the value of $Z_h$ and lead to a multi-graviton continuum $f_h$. For small spectral values, $f_h$ approaches a finite value which can be determined using perturbation theory in an effective theory below the Planck scale. For spectral values approaching the Planck scale and above, the spectral function becomes sensitive to the ultraviolet (UV) completion and non-perturbative techniques are required for its determination.  \smallskip

\textit{Spectral renormalisation group.---} 
To establish the existence of \cref{eq:rhoh-para}, we set up a functional renormalisation group (fRG) approach for Lorentzian quantum field theories, utilising the spectral functional framework developed in \cite{Horak:2020eng, Horak:2021pfr}. This approach is based on a modified dispersion $p^2 \to p^2 +R_k(p^2)$, where we use the Lorentz-invariant choice
\begin{align}
	\label{eq:CS-Cutoff}
	R_k = Z_\phi\, k^2\,.
\end{align}
This is a Callan-Symanzik (CS) cutoff including the on-shell wave-function renormalisation  $Z_\phi$ of the fluctuation fields $\phi=(h_{\mu\nu}, c_\mu,\bar c_\mu)$. The cutoff \cref{eq:CS-Cutoff} shifts the on-shell condition by $k^2$ to larger values without introducing poles or cuts into the propagator. Conversely, using a standard momentum-dependent Lorentz-invariant regulator $R_k(p^2)$ necessarily introduces poles and cuts in the complex plane. Then, \cref{eq:KS-L} does not hold at finite $k$. Hence, for the present study, we use \cref{eq:CS-Cutoff} which does not spoil \cref{eq:KS-L} from the outset.

While the cutoff \cref{eq:CS-Cutoff} is best suited to extract spectral data, it comes at a price: the corresponding fRG flow requires additional renormalisation because the standard UV divergences and counter terms resurface \cite{Symanzik:1970rt}. In practice,  local divergences of the flow must be absorbed in the parameters of the cutoff-dependent effective action. Here, we use dimensional regularisation, which respects the symmetries of the theory including gauge and diffeomorphism invariance, see \cite{Horak:2020eng, Horak:2021pfr}. This leads to a well-defined finite flow for effective actions $\Gamma_k$ with Euclidean or Lorentzian signature,
\begin{align}\label{eq:RenFlow}
	\partial_t\Gamma_k[\phi] =  \frac12 \Tr\, \CG_k[\phi]\,\partial_t \CR_k - \partial_t S_{\text{ct},k}[ \phi] \,.
\end{align}
Here, $\CR_k$ is the regulator matrix of all graviton and ghost modes. Similarly, $\CG_k[\phi] =1/({\Gamma_k^{(2)}[\phi]+ \CR_k})$ with $\Gamma_k^{(2)}\equiv\delta^2\Gamma_k/\delta\phi\delta\phi$ is the field-dependent propagator matrix at scale $k$, and we have introduced  the `RG time' parameter $t=\ln k/k_\text{ref}$ with a reference scale $k_\text{ref}$. 

The spectral flow \cref{eq:RenFlow} can be derived from the standard finite Wetterich flow \cite{Wetterich:1992yh} with spatial-momentum regulators $R_k({\vec p}^{\,2})\to Z_\phi k^2$ as briefly outlined in the supplement, see also  \cite{Braun:2022mgx}. Spatial-momentum regulators also preserve the spectral representation but break Lorentz invariance. The latter is restored in the above limit, in which also the counter terms $\partial_t S_{\text{ct},k}$ emerge naturally in a well-defined limit of finite flows. 

\begin{figure}[t!]
	\includegraphics[width=.95\linewidth]{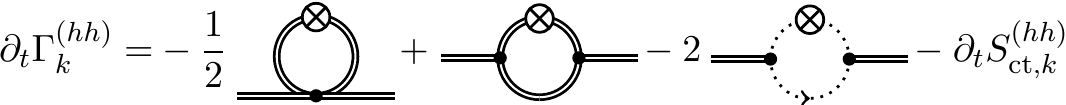}
	\caption{Flow of the graviton two-point function. Double (dotted) lines represent graviton (ghosts) propagators, dots indicate vertices, and the cross denotes a regulator insertion. 
	}
	\label{fig:FlowProp}
\end{figure}

With \cref{eq:RenFlow}, we can provide explicit flow equations for the graviton propagator or vertices. For example, the flow for the graviton two-point function follows from \cref{eq:RenFlow} through a vertex expansion of $\Gamma_k[\phi]$ about vanishing fluctuation field $\phi=0$. It is extracted from the graviton TT mode whose scalar propagator  reads $\CG_{hh} = (\Gamma^{(hh)}_\text{TT} + R_k)^{-1}$, with
\begin{align} \label{eq:Gamma2}
	\Gamma^{(hh)}_\text{TT}(p) = Z_h(p) ( p^2 +\mu\, k^2) \,.
\end{align}
Here, $Z_h(p)$ is the momentum-dependent graviton wave function, and $\mu$  the on-shell graviton mass parameter. With this parametrisation, the graviton propagator has a pole at $m_h^2= k^2(1+\mu)$, cf., the delta-peak in the spectral function \cref{eq:rhoh-para}. The wave-function renormalisation in \cref{eq:rhoh-para,eq:CS-Cutoff} is defined on-shell $Z_h\equiv Z_h(p^2=-m_h^2)$, the Lorentzian signature being key for this definition.

Schematically, the non-perturbative flow for the graviton two-point function is displayed in \cref{fig:FlowProp}. Apart from regulator insertions and prefactors, it resembles one-loop diagrams, though with non-perturbative propagators and vertices. We further need the flow of gravitational vertices, in particular the three-graviton vertex. Here, we limit ourselves to vertices at vanishing momentum, where we may exploit equations derived in Euclidean signature as these fall back onto their Lorentzian counterparts required here \cite{Christiansen:2015rva, Denz:2016qks}. Differences in the technical setup are subleading as long as the mass parameter stays away from off-shell poles, and the graviton anomalous dimension $\eta_h=-\partial_t \ln Z_h$ remains small.  \smallskip

\textit{Flow of the graviton spectral function.---}
We are now ready to provide an explicit non-perturbative flow for the graviton spectral function \cref{eq:rhoh-para}. Using the flow for the graviton propagator with \cref{eq:KS-L,eq:rho-G}, we find
\begin{align}\label{eq:dotrho-dotlambdahh}
	\partial_t \rho_h &= -2\,\text{Im}\, \CG_{hh}^2 \left(\partial_t \Gamma^{(hh)}_\text{TT} + \partial_t R_k\right),
\end{align}
where the right-hand side is evaluated at $p=-\imag (\lambda + \imag \varepsilon)$, and the present spectral approach allows us to take this limit analytically, see \cite{Horak:2020eng, Horak:2021pfr}. Using the spectral representation \cref{eq:KS-L} for gravitons and ghosts, all diagrams in \cref{fig:FlowProp} are now expressed as integrals over spectral values and a dimensionally regularised loop momentum. This reads 
\begin{align}\label{eq:dotIpol}
	\partial_t \Gamma^{(hh)}_\text{TT}\Big|_\text{3-point} =\prod_{i=1}^3  \int_0^\infty \frac{\dif\lambda_i^2}{2\pi} 
	\, \rho_h(\lambda_i) \ I_\text{3-point}(p,\{\lambda_j\}) \,,
\end{align}
for the diagram with graviton three-point vertices (second diagram in \cref{fig:FlowProp}), and similarly for the other diagrams. The three spectral values relate to the three propagators in the diagram, and the function $I_\text{3-point}$ accounts for all tensor contractions and a remaining loop momentum integration. The latter integral can be performed analytically.  In \cref{eq:dotIpol}, we only need the spectral function \cref{eq:rhoh-para} at $\vec p=0$  due to Lorentz invariance. For the single-graviton delta-peak, also the $\lambda_i$ integrals in \cref{eq:dotIpol} can be performed straightforwardly, leading to closed analytic flows.

The graviton spectral function is obtained by integrating the flow \cref{eq:dotrho-dotlambdahh}. Here, we solve \cref{eq:dotrho-dotlambdahh} without feeding back $f_h$ on the right-hand side. This contribution is subleading and will be considered elsewhere. \smallskip

\begin{figure}[t!]
	\centering
	\includegraphics[width=.9\linewidth]{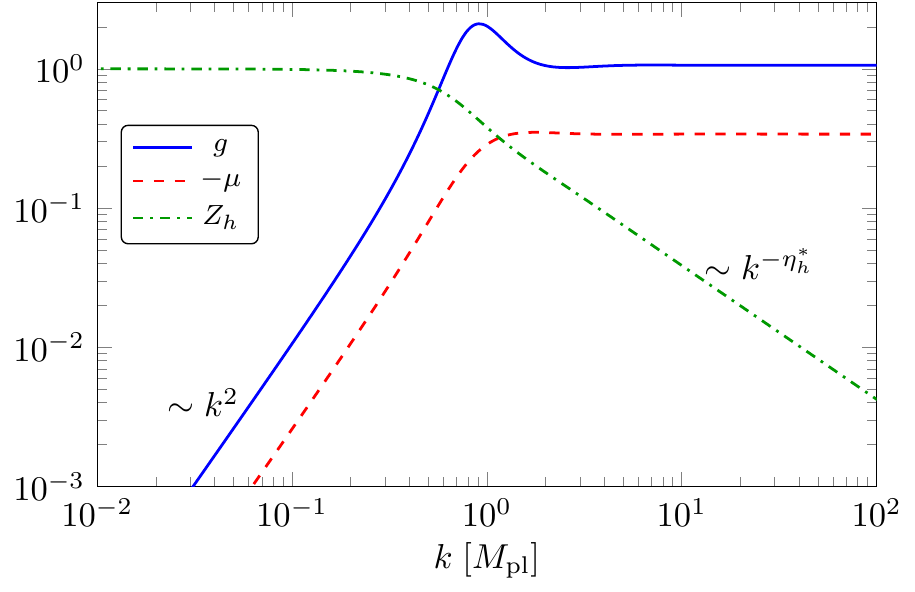}
	\caption{UV-IR connecting trajectory showing the dimensionless Newton coupling $g$, the graviton mass parameter $\mu$, and the graviton wave-function renormalisation $Z_h$.}
	\label{fig:trajectories}
\end{figure}

\textit{Single-graviton peak.---}
We start with the flow of the single-graviton delta-peak. Remarkably, our on-shell flows do not suffer from poles in the graviton propagator   $(\mu=-1)$ which are commonplace in off-shell studies. The three-graviton vertex, evaluated at vanishing momentum, provides the flow for Newton's coupling $G_\text{N}(k)= g(k)/k^2$ with an asymptotically safe UV fixed point 
\begin{align} \label{eq:FP}
	(g,\, \eta_h,\,\mu)\big|_*=(1.06,\,0.96,\,-0.34)\,.
\end{align}
The scaling exponents $\theta= 2.49 \pm 3.17\,\imag$ compare well with those found in Euclidean studies. To connect the short-distance fixed point \cref{eq:FP} with general relativity \cref{eq:EH-Action} at large distances, we impose the boundary conditions 
\begin{align}\label{boundary}
	(G_\text{N}(k),Z_h(k),k^2\mu(k))\big|_{k\to0}=(G_\text{N},1,-2\Lambda)\,,
\end{align}
where we have identified the infrared (IR) mass term with the cosmological constant in \cref{eq:EH-Action}. Note that for normalisable spectral functions with $\int\!\lambda \rho(\lambda) \text{d}\lambda=1$, the on-shell value of the wave function follows from this normalisation. The on-shell choice $Z_h=1$ is only possible as $\rho_h$ cannot be normalised: $\int\! \lambda \rho_h(\lambda)\text{d} \lambda=\infty$ following from its scaling in the UV regime, see \cite{Bonanno:2021squ}.

For now, we demand $\Lambda$ to vanish. Besides being viable phenomenologically, it also ensures that the on-shell condition on a flat Minkowski background remains satisfied. The resulting RG trajectory for $(g,Z_h,\mu)$ is displayed in \cref{fig:trajectories}, with the Planck scale set to $M^2_\textrm{pl}=1/G_\text{N}$. We observe that $Z_h$ becomes a constant in the IR while it scales as $\sim k^{\eta_h}$ in the UV, whereas $g$ and $-\mu$ scale $\sim k^2$ in the IR and settle at fixed points in the UV. The spike for $g$ around the Planck scale can be traced back to the complex conjugate nature of the scaling exponents.
\smallskip

\begin{figure}[t!]
	\centering
	\includegraphics[width=0.96\linewidth]{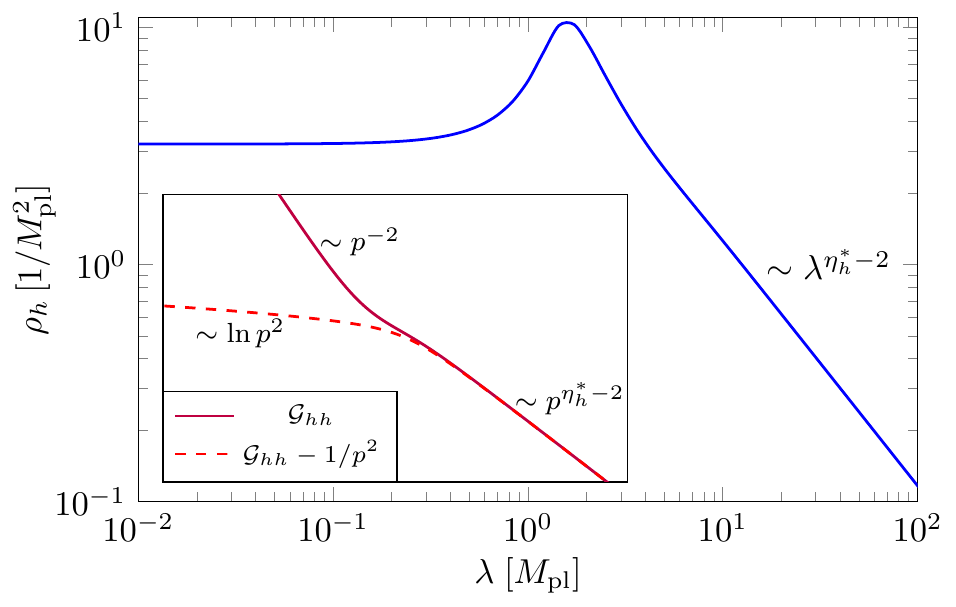}
	\caption{The graviton spectral function. The inset shows the reconstructed Euclidean propagator (full line) and the subleading logarithm (dashed). 
	}    
	\label{fig:rho_h-best}
\end{figure}

\textit{Multi-graviton continuum.---}
The multi-graviton continuum is found by integrating the flow \cref{eq:dotrho-dotlambdahh} with \cref{eq:rhoh-para} on the trajectory  displayed in \cref{fig:trajectories}. Structurally, the flow is proportional to $\theta(\lambda^2-4 m_h^2)$ with the largest contribution at the threshold. Consequently, the spectral function at $\lambda$ is predominantly built from quantum fluctuations at $k\approx \lambda/(2\sqrt{1+\mu})$ which supports our approximation of dropping the multi-graviton continuum $f_h$ on the right-hand side of the flow. Our result for $f_h$ is shown in \cref{fig:rho_h-best}. The function $f_h$ approaches a constant below the Planck scale, and scales as $\sim \lambda^{\eta_h^*-2}$ above the Planck scale. The spike near the Planck scale can be traced back to the complex conjugate scaling exponents, as was the case for $g$. Overall, the spectral function contains a massless $\delta$-peak and a positive multi-graviton continuum, constant in the IR and with an asymptotically safe scaling in the UV. The same attributes were found in the recent reconstruction from Euclidean data \cite{Bonanno:2021squ}.

\begin{figure*}

	\includegraphics[width=.38\linewidth]{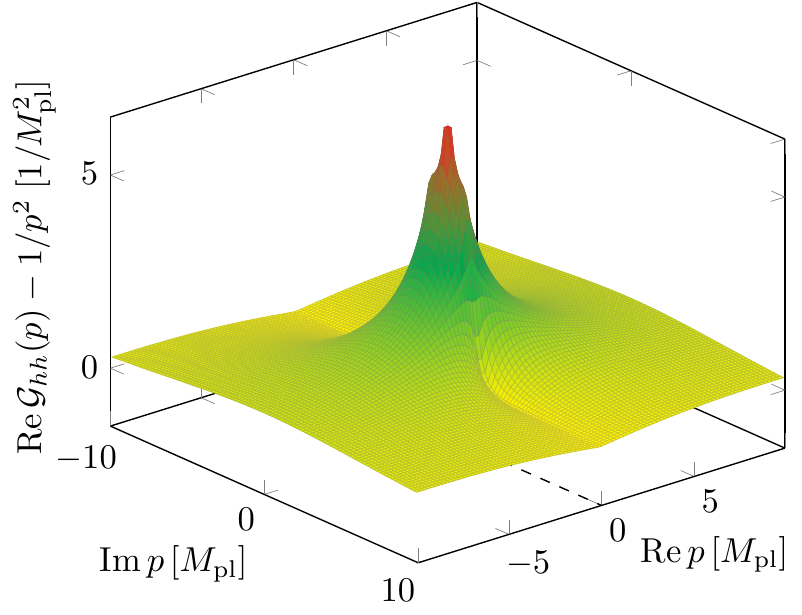}
	\qquad \qquad \qquad 
	\includegraphics[width=.38\linewidth]{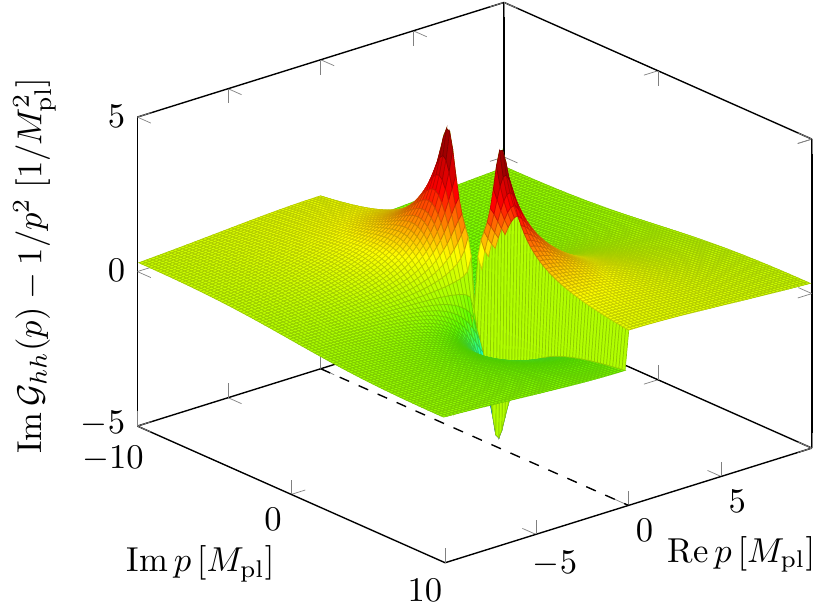}
	\caption{Real and imaginary part of the graviton propagator in the complex plane. The dashed line indicates the timelike axis.
	}
	\label{fig:3d-prop}
\end{figure*}

The finite value of the spectral function in the IR implies the presence of a subleading logarithm in the propagator $\CG_{hh}\sim p^{-2} -A_h \ln p^2+\,$subleading, as highlighted in the inset in \cref{fig:rho_h-best}. The coefficient $A_h$ is universal (regulator-independent) but gauge-dependent \cite{Kallosh:1978wt, Bonanno:2021squ}. It can be determined within effective theory, giving $A_h=61/(60\pi)\approx 0.32$. Conversely, integrating the  flow gives $A_h = 35/(9\sqrt{3})-11/(2\pi)\approx 0.49$. The difference is due to the neglected feedback of $f_h$, and serves as an indicator for subleading corrections.  We conclude that our approximation does not affect the leading behaviour of the propagator or global characteristics of the spectral function. We remark that the gauge dependence of the spectral function, which can be computed exactly in the IR via effective theory, is also present in the UV. Only the on-shell graviton $\delta$-peak is gauge-independent.

With the spectral function and using \cref{eq:KS-L}, we have access to the propagator in the whole complex momentum plane. The real and imaginary parts of the propagator are depicted in \cref{fig:3d-prop}, where we excluded the pole contribution in the real part.  Both parts vanish for asymptotically large $p$. The real part displays a unique pole at vanishing $p$ (not shown in \cref{fig:3d-prop}), while the imaginary part shows a branch cut along the timelike axis. \smallskip

\textit{Cosmological constant.---}
Next, we turn to Lorentzian quantum gravity with a non-vanishing cosmological constant. On de~Sitter (dS) or anti-de~Sitter (AdS) backgrounds, the classical graviton and ghost continue to be massless, and graviton vertices are deformed in comparison with flat backgrounds. Since alterations of the geometry are relevant for large spatial distances, we expect to find modifications of the spectral function at small spectral values. We continue to use flat backgrounds as above, meaning that our setup at $\Lambda\neq 0$ becomes an off-shell expansion. For simplified trajectories
\begin{align}
	\label{eq:gk-simple}
	G_\text{N}(k) = \frac{g^*}{k^2+g^* M_\text{pl}^2}\,,
\end{align}
the spectral flows admit analytic solutions which facilitate the present qualitative discussion. In \cref{eq:gk-simple}, $g^*$ takes the role of a free parameter. Furthermore, we neglect the ghost contributions. The respective UV fixed point is given by
\begin{align}
	\label{eq:FP-mu-simple}
	\mu^*&=\frac{-g^* }{c_\mu  + g^*}\,,
	&
	\eta_{h}^*&=\frac{2 g^* }{2 c_\eta  + g^*}\,,
\end{align}
with $(c_\mu,\, c_\eta) = ( 1.77,\, 0.49)$ known analytically and provided in the supplement. Using $g^*= 1.06$ from \cref{eq:FP}, we find $\mu^*= -0.38$ and $\eta_h^* =1.04$, both values being approximately 10\% off, see \cref{eq:FP}. This indicates that the ghost contributions are indeed subleading. 

The flow is readily integrated analytically with the IR boundary conditions \cref{boundary},
\begin{align}
	\label{eq:mu-sol}
	Z_{h}(k) &= \left( 1 + \frac{1}{c_\eta\,\eta_h^*} \frac{k^2}{M_\text{pl}^2} \right)^{\!\!-\frac12 {\eta_h^*}}, \notag  \\[2ex]
	\mu(k)&= \mu^* -\frac{2 \Lambda}{k^2} + \frac{c_1 M_\text{pl}^2-2\Lambda}{k^2} \left[Z_h(k)^{- c_2}-1\right],
\end{align}
with  $c_1=2.17\, g^* /(1.77  + g^*)$ and $c_2 =0.45$ (further details including analytical expressions are given in the supplement). 

\begin{figure}[b]
	\includegraphics[width=.9\linewidth]{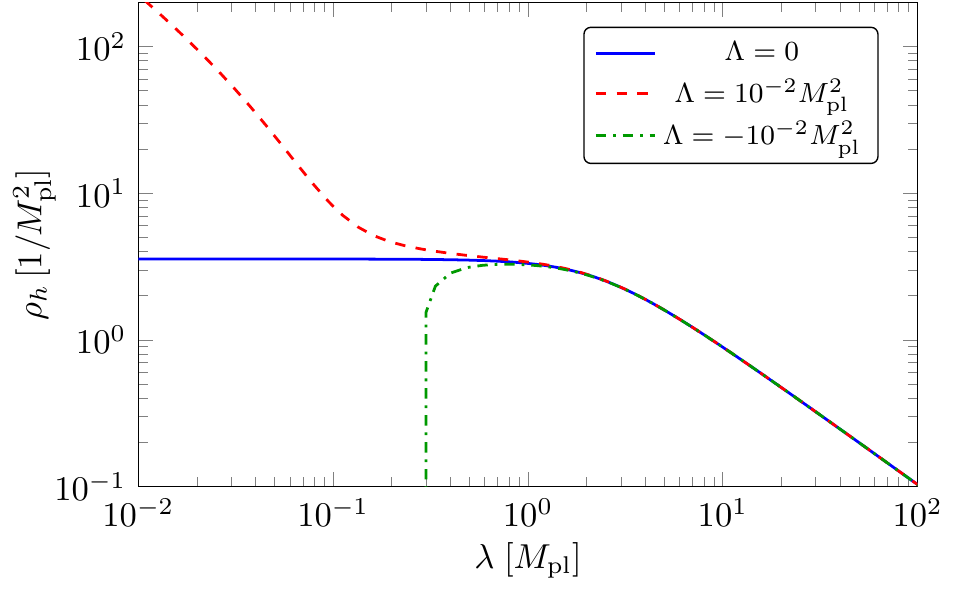}
	\caption{Enhancement (or suppression) of the  spectral function due to a positive (or negative) cosmological constant.
	}
	\label{fig:finite-Lambda}
\end{figure}

Several comments are in order. For $g^*$ taking real positive values, the graviton anomalous dimension ranges within $\eta_h^* \in (0,2)$. We therefore have  $Z_h\to 1$ in the IR, and $Z_h\to 0$ in the UV with a power-law that mildly depends on  $g^*$, reminiscent of the full solution for $\Lambda=0$ (\cref{fig:trajectories}). The crossover sets in at $k^2/M_\text{pl}^2\approx c_\eta \eta^*_h $ which is close to but smaller than the Planck scale. Remarkably, the short distance mass parameter is constrained within the narrow range $\mu^*\in (-1,0)$ and only takes negative values. From the explicit result \cref{eq:mu-sol}, and also observing $c_2\eta^*_h<1$, it is evident that the mass parameter $\mu(k)$ interpolates smoothly between $\mu^*$ in the UV  and the cosmological constant $-2\Lambda/k^2$ in the IR. We conclude that \cref{eq:gk-simple,eq:mu-sol} are viable approximate solutions interpolating between an asymptotically safe fixed point and general relativity with a cosmological constant.

Following the same steps as before, we can now find the spectral function for $\Lambda\neq 0$ by integrating the flow \cref{eq:dotrho-dotlambdahh} with \cref{eq:rhoh-para} along the trajectories \cref{eq:gk-simple,eq:mu-sol}. Our results are illustrated in \cref{fig:finite-Lambda}. We observe that a positive or negative cosmological constant does not affect the spectral function for spectral values above $\lambda \gtrsim \sqrt{|8\Lambda|}$. For smaller spectral values, the geometry leaves an imprint. For AdS backgrounds, the cosmological constant acts like a mass term which leads to a suppression. Conversely, the spectral function is enhanced for dS backgrounds because $\Lambda>0$ acts like a negative mass-squared term.

The off-shell effects due to the cosmological constant become even more pronounced if the ghost contributions are retained. The ghost remains on-shell at $k^2$ compared to the off-shell graviton at $m_h^2=k^2(1+\mu)$. We find that for AdS backgrounds (at $\mu= 3$), off-shell gravitons can directly scatter into the on-shell multi-ghost continuum and the flow of $f_h$ diverges, while it stays finite for dS backgrounds. In this off-shell computation, the flat Minkowski background bears similarities to an \textit{external} electric or magnetic field in QED. External backgrounds or boundary conditions can introduce driving forces or friction that constantly feed or suppress scattering processes, which then destroy unitarity much like in open quantum systems. This analogy allows for a heuristic interpretation of the AdS singularity in the flow: there the off-shell background serves as a driving force for graviton scattering processes. We expect that full on-shell AdS flows with ghost contributions remain finite. Then, graviton and ghost are both on-shell massless, and it is the off-shell shift of mass scales that triggers the divergence. \smallskip

\textit{Discussion \& Conclusion.---}
We have put forward the first direct computation of the graviton spectral function in quantum gravity. The spectral function shows a massless one-graviton peak and a positive multi-graviton scattering continuum  (\cref{fig:rho_h-best}), interpolating between a constant part for small and an asymptotically safe scaling regime for large spectral values. While the spectral function can always be defined as the imaginary part of the retarded propagator \cref{eq:rho-G}, the KL representation \cref{eq:KS-L} only holds if the propagator has no poles or cuts in the complex upper half plane. Therefore, it is quite remarkable that the graviton spectral function and propagator \textit{indeed} obey the KL representation \cref{eq:KS-L} with a positive spectral function and no ghost or tachyonic instabilities. The absence of the latter instabilities is crucial for the unitarity of the theory. This noteworthy result should be contrasted with the unclear situation in non-Abelian gauge theories where a similar understanding has not yet been achieved~\cite{Cyrol:2018xeq, Li:2019hyv, Binosi:2019ecz, Hayashi:2021nnj, Hayashi:2021jju, Horak:2021pfr, Kluth:2022wgh}.

On the technical side, and to ensure that the KL representation \cref{eq:KS-L} is not inadvertently spoiled by the momentum cutoff,  the spectral flow necessitates \textit{spectral} regulators which do not introduce cuts and poles in the complex upper half plane. In our study, we have explicitly observed the absence of the latter, which therefore guarantees a spectral representation  for all scales. Further, we have advocated the unique Lorentz-invariant \textit{spectral} cutoff \cref{eq:CS-Cutoff}, at the expense of an additional regularisation \eqref{eq:RenFlow}. The latter can be avoided by using spatial cutoffs, though at the price of breaking Lorentz invariance.  Still, the corresponding flows are linked to the CS spectral flow in well-defined limits, and offer avenues for systematic error estimates.

Finally, we note that our findings open a door to investigate scattering amplitudes and unitarity of fully quantised  gravity~\cite{Denz:2016qks, Bonanno:2021squ, Knorr:2019atm, Draper:2020bop, Draper:2020knh, Platania:2020knd}. The key building blocks are the timelike graviton propagator obtained here (\cref{fig:3d-prop}), and the corresponding spectral functions for scattering vertices. Extracting vertices from \cref{eq:RenFlow,eq:dotrho-dotlambdahh} is in reach, albeit technically more demanding than extracting propagators. We thus look forward to direct tests of unitarity in asymptotically safe gravity. \smallskip

\textit{Acknowledgements.---}
We thank A.~Bonanno, T.~Denz, J.~Horak, B.~Knorr, J.~Papavassiliou, A.~Platania, and N.~Wink for discussions. This work is funded  by Germany’s Excellence Strategy EXC 2181/1 - 390900948 (the Heidelberg STRUCTURES Excellence Cluster) and the DFG Collaborative Research Centre "SFB 1225 (ISOQUANT)", and is supported by the Science and Technology Research Council (STFC) under the Consolidated Grant ST/T00102X/1.

\bibliographystyle{apsrev4-1}
\bibliography{GravityStatus}


\clearpage

\renewcommand{\thesubsection}{{S.\arabic{subsection}}}
\setcounter{section}{0}

\onecolumngrid

\section*{Supplemental material}

In this supplement, we provide technical details omitted in the main text. In \cref{sec:Gauge}, we detail the gauge-fixing and ghost action, while \cref{sec:Prop+Vertices} provides the transverse-traceless projection of the graviton. In \cref{sec:Diagrams}, we provide the relevant expressions for the evaluation of loop diagrams. In \cref{sec:DerRenCS}, we briefly outline the derivation of the flow equation \cref{eq:RenFlow}.  In \cref{sec:flows}, we discuss renormalised flows in the presence of a Callan-Symanzik cutoff. In \cref{sec:complex-plane}, we provide further details for the propagator in the complex plane. In \cref{sec:parameters}, we offer details for the derivation of analytical solutions and for the expressions \cref{eq:FP-mu-simple,eq:mu-sol} stated in the main text.

The computations were performed using the Mathematica platform and an array of additional libraries: VertEXpand \cite{vertexpand} and DoFun \cite{Huber:2019dkb, github:DoFun} depending on \cite{xPerm, Brizuela:2008ra}, FormTracer \cite{Cyrol:2016zqb, github:FormTracer} depending on \cite{Feng:2012tk, Ruijl:2017dtg}, and HypExp \cite{Huber:2005yg}.

\subsection{Gauge-fixing and ghost action}\label{sec:Gauge}

The Einstein-Hilbert action \cref{eq:EH-Action} is augmented by a de-Donder type gauge-fixing, 
\begin{align}
	\label{eq:Sgf}
	S_\text{gf}[\bar g,h] &= \frac1{2\alpha}\int \! \mathrm d^4 x\, \sqrt{\bar g}\,\bar g^{\mu\nu}F_\mu F_\nu \,, 
	&&\text{with}&
	F_\mu &= \bar \nabla^\nu h_{\mu\nu} - \frac{1+\beta}4 \bar \nabla_\mu {h^\nu}_\nu \,,
\end{align}
and with the respective ghost action 
\begin{align}
	S_\text{gh}[\bar g, h, \bar c, c] &= \int \! \mathrm  d^4 x \,\sqrt{\bar{g}}\, \bar{c}^\mu \mathcal{M}_{\mu\nu} c^\nu \,,
	&&\text{with}&	
	\mathcal{M}_{\mu\nu} &= \bar\nabla^\rho\left(g_{\mu\nu}\nabla_\rho+g_{\rho\nu}\nabla_\mu \right) - \frac{1+\beta}{2} \bar{g}^{\rho \sigma} \bar{\nabla}_\mu\left( g_{\nu \rho} \nabla_\sigma \right) .
\end{align}
The Faddeev-Popov operator $\mathcal{M}$ follows from a diffeomorphism variation of the gauge-fixing condition \cref{eq:Sgf}. The ghost spectral function $\rho_c$ is parametrised in analogy to the graviton spectral function \cref{eq:rhoh-para} with the replacements $m_h\to k$, $Z_h\to Z_c$, and $f_h\to f_c$, $Z_c$ being the on-shell ghost wave-function renormalisation. 

Throughout this work, we use the harmonic gauge $\alpha=\beta=1$. We remark that the Landau limit $\alpha\to0$ introduces non-localities in the diagrams, leading to terms $p^4 \log p^2 $ in $\partial_t \Gamma^{(hh)}$, see also \cite{Knorr:2021niv}. This is related to the fact that the loop integrals involve projection operators $\Pi_\text{TT}(q)$ as well as $\Pi_\text{TT}(p+q)$. While they vanish at $p=0$, they obstruct the analytic continuation.

\subsection{Propagator and Vertices}\label{sec:Prop+Vertices}

In this work, we focus on the correlation functions of transverse-traceless gravitons. The transverse-traceless tensor structure $\Pi_\text{TT}(p)$ is given by 
\begin{align}\label{eq:Pi_TT} 
	\Pi_\text{TT}^{\mu\nu \rho\sigma}(p) &= \Pi^{\mu(\rho}(p) \Pi^{\sigma)\nu}(p)  - \frac{1}{3} \Pi^{\mu\nu}(p) \Pi^{\rho\sigma}(p)\,,
	&&\text{with}&
	\Pi^{\mu\nu}(p) &= \eta^{\mu\nu} - \frac{p^\mu p^\nu}{p^2}\,,
\end{align}
where the parenthesis in the superscript stand for symmetrisation with respect to the indices $\rho$ and $\sigma$: $O_1^{(\rho}\,O_2^{\sigma)}=1/2 \,( O_1^{\rho}\,O_2^{\sigma}+O_1^{\sigma}\,O_2^{\rho})$. The subtraction in \cref{eq:Pi_TT} leads to $(\Pi_\text{TT})^\mu{}_\mu{}^\rho{}_\rho=0$, and we have $\Pi_\text{TT}^2= \Pi_\text{TT}$. The graviton two-point function has the parametrisation, 
\begin{align}\label{eq:Gahh}
	\Gamma^{(hh),\mu\nu\rho\sigma}&= \Gamma^{(hh)}_\text{TT}\Pi^{\mu\nu\rho\sigma}_\text{TT} +\text{other modes}\,,
	&&\text{with}&
	\Gamma^{(hh)}_\text{TT} &= Z_h(p^2) \left( p^2 +\mu\, k^2\right) \,,
\end{align}
c.f., \cref{eq:Gamma2}. In \cref{eq:Gahh} we have dropped the $\delta$-function which guarantees momentum conservation. The respective transverse-traceless graviton propagator is given by
\begin{align}\label{eq:GTT}
	\CG^{\mu\nu\rho\sigma}_{hh,\text{TT}}(p) &= \CG_{hh}(p) \Pi_\text{TT}^{\mu\nu\rho\sigma}(p)\,,
	&&\text{with}&
	\CG_{hh}(p) &=\frac{1}{\Gamma^{(hh)}_\text{TT} + Z_h k^2}\,.
\end{align}
For the scalar propagator function $\CG_{hh}(p)$, we use the KL spectral representation, c.f.~\cref{eq:KS-L}. We describe all other modes of the graviton propagator by the same uniform scalar propagator function.

In the flow of the propagator in \cref{fig:FlowProp}, we are using the classical $n$-graviton vertices derived from $n$ metric-derivatives of the Einstein-Hilbert action \cref{eq:EH-Action} with vanishing cosmological constant. The approximation of a vanishing cosmological constant in the vertices is supported by the Euclidean results in \cite{Denz:2016qks}. These classical vertices are dressed with the on-shell graviton wave-function renormalisation, which takes care of the renormalisation properties of the graviton legs,
\begin{align}\label{eq:Vertapprox}
	\Gamma^{(h_1\cdots h_n)}(p_1,\ldots,p_n)= Z^{n/2}_h \, S_\text{EH}^{(h_1\cdots h_n)}(p_1,\ldots,p_n)\Big|_{\Lambda\to0}\,,
\end{align}
and analogously for the ghost-graviton vertices. We emphasise that $Z_h\equiv Z_h(p^2=-m_h^2)$ is the on-shell wave-function renormalisation. Note that the metric split $g_{\mu\nu}= \eta_{\mu\nu}+\sqrt{16 \pi G_\text{N}} \,h_{\mu\nu}$ makes the propagator independent of the Newton coupling $G_\text{N}$.

\subsection{Evaluation of loop diagrams}\label{sec:Diagrams}

There are three diagrams contributing to the flow of the graviton two-point function, $\partial_t \Gamma^{(hh)}_\text{TT} =  \partial_t \Gamma^{(hh)}_\text{TT}|_\text{tadpole}+ \partial_t \Gamma^{(hh)}_\text{TT}|_\text{3-point}+ \partial_t \Gamma^{(hh)}_\text{TT}|_\text{ghost}$, see \cref{fig:FlowProp}. After using the KL spectral representation \cref{eq:KS-L}, they read schematically
\begin{align} \label{eq:flow-Gamma2}
	\partial_t \Gamma^{(hh)}_\text{TT}\Big|_\text{tadpole} =&\,\prod_{i=1}^2  \int_0^\infty \frac{\dif\lambda_i}{\pi}\lambda_i\, \rho_h(\lambda_i) \int \frac{\dif^d q}{(2\pi)^d}  \frac{V_{\text{tadpole}}(p, q)}{(q^2+\lambda_1^2) (q^2+\lambda_2^2)} \,, \notag \\[1ex]
	\partial_t \Gamma^{(hh)}_\text{TT}\Big|_\text{3-point} =&\,\prod_{i=1}^3  \int_0^\infty \frac{\dif\lambda_i}{\pi}\lambda_i\, \rho_h(\lambda_i)\int\! \frac{\mathrm d^d q}{(2\pi)^d} \frac{V_\text{3-point}(p, q)}{(q^2+\lambda_1^2) (q^2+\lambda_2^2)\left((p+q)^2+\lambda_3^2\right)} \,, \notag \\[1ex]
	\partial_t \Gamma^{(hh)}_\text{TT}\Big|_\text{ghost} =&\,\prod_{i=1}^3  \int_0^\infty \frac{\dif\lambda_i}{\pi}\lambda_i\, \rho_c(\lambda_i) \int\! \frac{\mathrm d^d q}{(2\pi)^d} \frac{V_\text{ghost}(p, q)}{(q^2+\lambda_1^2) (q^2+\lambda_2^2)\left((p+q)^2+\lambda_3^2\right)}\,.
\end{align}
The second line in \cref{eq:flow-Gamma2} is schematically the same as \cref{eq:dotIpol} in the main text. The factors $V_i$ combine the contractions of the vertices with the regulator derivative $\partial_t R_k = (2 -\eta_h) k^2$. With the abbreviation $s = (p+q)^2$, they read
\begin{align} \label{eq:V-factors}
	V_\text{tadpole}=&\, -  16\pi g  \left( 2- \eta_h \right) (4 p^2 + 3 q^2) \,, \notag \\[1ex]
	V_\text{3-point} =&\,	 \frac{8\pi g ( 2 - \eta_h  ) }{15 p^4}  (71 p^8 + (q^2 - s)^4 + 16 p^6 (q^2 + s) + 6 p^2 (q^2 - s)^2 (q^2+ s) + p^4 (26 q^4 + 4 q^2 s + 26 s^2))\,,\notag \\[1ex]
	V_\text{ghost} =&\, - \frac{40\pi g ( 2 - \eta_c )}{3} (p^4 + q^4 + 10 q^2 s + s^2 - 2 p^2 (q^2 + s)) \,.
\end{align}
In summary, this leads us to momentum integrals of the type
\begin{align}
	T_{\alpha\beta\gamma} &= \int \! \frac{\dif^d q}{(2\pi)^d} \frac{p^{2 \alpha} q^{2\beta} s^{\gamma}}{(q^2+\lambda_1^2) (q^2+\lambda_2^2) (s+\lambda_3^2)} \,,
\end{align}
in $d=4-2\epsilon$ dimensions for $\alpha = -2,\ldots,2$, and $\beta,\gamma = 0,\ldots,4$. The integral is conveniently rewritten in a symmetrised version with respect to $\lambda_1$ and $\lambda_2$ as
\begin{align}
	T_{\alpha\beta\gamma} &=  \frac{p^{2 \alpha}}{\lambda_2^2-\lambda_1^2} \cdot \tilde{T}_{\beta\gamma} + (\lambda_1 \leftrightarrow \lambda_2)	\,,
	&
	\tilde{T}_{\beta\gamma} &=  \int \! \frac{\dif^d q}{(2\pi)^d} \frac{q^{2\beta} s^{\gamma}}{(q^2+\lambda_1^2) (s+\lambda_3^2)} \,.
\end{align}
This is a one-loop integral with propagators of massive fields that can be solved with standard methods. The resulting expressions are too long to be displayed here but can be found in a supplemented Mathematica notebook.

\subsection{Derivation of renormalised spectral Callan-Symanzik flow}\label{sec:DerRenCS}

The \textit{finite} spectral flow equation \labelcref{eq:RenFlow} with flowing renormalisation can be derived from standard \textit{finite} momentum cutoff flows by considering both UV and IR momentum cutoffs in a combined regulator function 
\begin{align}
	 R_{k,\Lambda(k)}(p)\,,
\label{eq:IR+UVR} 
	\end{align}
where $\Lambda(k)$ indicates a change of the UV cutoff that accompanies that of the IR scale $k$. The regulator \labelcref{eq:IR+UVR} has to be a combination of IR regulators $R_{k,\Lambda(k)}$, that also incorporate a UV regularisation. The latter is then removed in a controlled way, leading to the emergent flow of the local counter-term action. For example, the regulator
\begin{align}
	R_{k,\Lambda(k)} =  Z_\phi\, k^2 \left( \theta(1 - \vec p^{\,2}/ \Lambda_1(k)^2)+ \left[\frac{1}{\theta(1-\vec p^{\,2}/ \Lambda_2(k)^2)}-1\right]\right),
\label{eq:example-reg} 
\end{align}
which is a combination of a flat cutoff \cite{Litim:2000ci,Litim:2001up} and a sharp cutoff, can be used to remove one divergence and fix one renormalisation condition for suitable choices of $\Lambda_1(k)$ and $\Lambda_2(k)$. In case of more divergences, the respective regulator function is more complicated, but it was shown in \cite{Braun:2022mgx} that suitable combinations of regulators always exist. This allows us to take the limit
\begin{align}
	\lim_{\Lambda(k)\to \infty} R_{k,\Lambda(k)}(p) = Z_\phi \,k^2\,,
\label{eq:CS-limitk-Lambda}
\end{align}
in a manifestly finite way. The induced change of the UV cutoff scale implies a change of the UV renormalisation, called \textit{flowing renormalisation}. The details of the derivation of functional renormalisation group equations with flowing renormalisation, and specifically that of the renormalised CS equation put forward here can be found in~\cite{Braun:2022mgx}. The derivation provides that the flow equation \labelcref{eq:RenFlow} can be used with a CS cutoff in dimensional regularisation, and without explicitly taking the limit \labelcref{eq:CS-limitk-Lambda}.

Here, we briefly outline the main technical idea that follows from the UV-IR cutoff setup introduced above. Taking a $t$-derivative of the $k$- and $\Lambda$-dependent effective action leads us to the combined flow  
\begin{align}
\bigl( 	\partial_t+\left[\partial_t\Lambda(k) \right]\,\partial_\Lambda \bigr) \Gamma_{k,\Lambda}[\phi]= \frac12 {\rm Tr} \; {\cal G}_{k,\Lambda} \bigl( \partial_t  {\cal R}_{k,\Lambda}+\left[\partial_t\Lambda(k) \right]\, \partial_\Lambda {\cal R}_{k,\Lambda}\bigr),
	\label{eq:t-LambdaFlow}
\end{align}
where the subscripts ${}_{k,\Lambda}$ indicate the presence of the IR and UV momentum cutoffs. As $\partial_t\Lambda(k)$ is up to our disposal, we can choose it for the implementation of specific renormalisation conditions in the theory at hand. The above flow is manifestly finite. Moreover, the UV cutoff $\Lambda(k)$ can  be chosen such that a given set of renormalisation conditions for the effective action $\Gamma_{k,\Lambda(k)}$ is either kept unchanged or is changed in a specific way. This procedure is called \textit{flowing renormalisation}. The choices of the regulator function and $\Lambda(k)$ do not influence the results shown here, as long as the CS cutoff is approached in the limit $\Lambda(k)\to \infty$, where the flow of the local counter-term action emerges and can be computed within dimensional regularisation. In fact, it is sufficient to start with a CS cutoff and suitable renormalisation conditions from the outset as done here. Naturally, the respective term in \labelcref{eq:t-LambdaFlow} should be understood as a \textit{generalised} counter-term action, and we define 
\begin{align}\label{eq:predeltaFlowScl}
	\partial_t S_{\textrm{ct},\Lambda}[\phi]:= - \frac12 {\rm Tr} \, {\cal G}_{k,\Lambda}\, \left[\partial_t\Lambda(k) \right]\, \partial_\Lambda {\cal R}_{k,\Lambda}\,.  
\end{align}
In contrast to standard counter-term actions,  $S_{\textrm{ct},\Lambda}[\phi]$ in \labelcref{eq:predeltaFlowScl} is not local in general: its definition in terms of a one-loop flow comprises all powers in the fields as well as non-polynomial momentum dependences for finite $\Lambda$. A local version is approached in the limit $\Lambda\to \infty$. Then, all terms can be ordered in powers of $\Lambda$ and only positive powers and logarithms survive while maintaining the manifest finiteness of \labelcref{eq:t-LambdaFlow}. This leads us to the finite flow 
\begin{align}\label{eq:RenFlowSup}
	\partial_t\Gamma_k[\phi] =  \frac12 \Tr\, \CG_k[\phi]\,\partial_t \CR_k - \partial_t S_{\text{ct},k}[ \phi] \,,
\end{align}
with 
\begin{align}\label{eq:deltaFlowScl}
	\partial_t S_\textrm{ct}[\phi]=  \lim_{\Lambda\to\infty}	\partial_t S_{\textrm{ct},\Lambda}[\phi]\,.
\end{align}
We emphasise that \labelcref{eq:deltaFlowScl} is a formal definition, in general neither the first nor the second term in \labelcref{eq:RenFlowSup} are separately finite in the limit $\Lambda\to \infty$, but the combination is. The explicit results for $\partial_t S_\text{ct}$ are provided in the next supplement. More details on the local limit \labelcref{eq:RenFlowSup} with \labelcref{eq:deltaFlowScl} can be found in \cite{Braun:2022mgx}.  For a discussion of other real-time fRG approaches see e.g.\ \cite{Dupuis:2020fhh}.

\subsection{Flow equations and renormalisation}\label{sec:flows}

The flow of the graviton two-point function stems from three diagrams, see \cref{eq:flow-Gamma2,eq:V-factors} as well as \cref{fig:FlowProp}. These flows still contain $1/\varepsilon$-divergences that need to be renormalised. In comparison to perturbation theory, the degree of divergence is reduced due to the cutoff line which contains an additional propagator. Thus, for Einstein-Hilbert propagators, it has an additional decay with $1/p^2$ for large momenta. In the standard Euclidean fRG approach with a sufficiently fast decaying regulator, this additional propagator is irrelevant for the convergence properties of the loops. The CS-cutoff does not decay with momenta, so the degree of divergence of the diagrams is reduced by $-2$ in comparison to perturbation theory, and the flowing counter term action $\partial_t S_\textrm{ct}[\phi]$ is leading to a finite flow, see supplement \labelcref{sec:DerRenCS} for a brief discussion and for an in-detail derivation see  \cite{Braun:2022mgx}. 

Due to the reduction of the degree of divergence by $-2$, the CS-equation for gravity has at most quadratic divergences instead of the quartic ones of perturbation theory. Moreover, all terms with logarithmic divergences in perturbation theory are finite in the CS-equation. In summary, the CS-fRG has two divergences: 
\begin{itemize} 
	\item[(i)] graviton mass parameter $\mu$: quadratic divergence 
	\item[(ii)] wave function $Z_h(p=0)$: logarithmic divergence 
\end{itemize}
and hence 
\begin{align}\label{eq:CTs}
	\partial_t S^{(hh)}_{\text{ct},\text{TT},k}[\eta,0](p) = \left( c_1\, p^2 + c_0\,k^2\right) \Pi_\text{TT}(p)\,.
\end{align}
Importantly, all $p^4$-terms are finite. The loop integrals in \cref{fig:FlowProp} are carried out in $d=4-2\varepsilon$ dimensions and we parametrise the coefficients in \cref{eq:CTs} with $c_i =  \frac{c_{i,0}}{\varepsilon} + c_{i,1}$. The $1/\varepsilon$ terms compensate the divergences of the loops, while the finite parts are fixed by our choice of renormalisation conditions. In this work, we chose a renormalisation at vanishing momentum, $\partial_t \Gamma^{(hh)}(p=0)=0$ and $\partial_t \partial_{p^2} \Gamma^{(hh)}(p=0)=0$, which implies with the parametrisation \cref{eq:Gamma2},
\begin{align} \label{eq:RG-conditions}
	\partial_t \left(Z_h(p=0) \mu k^2\right) =&\,0\,, 
	&
	\partial_t \left(Z_h(p=0) + \mu k^2 \partial_{p^2} Z_h(p=0) \right) =&\,0\,.
\end{align}
Beyond the present approximation it is suggestive to choose an 'on-shell' renormalisation at $p^2=\mu k^2$ for all cut-off scales, and also compute the Newton constant at this momentum scale. This interesting extension goes beyond the scope of the present work and will be discussed elsewhere. 

The structure of \cref{eq:CTs} carries over to all $n$-point functions: their flows carry a quadratic divergence in the constant term and a logarithmic one in the $p^2$ one. In the physical limit, $k\to 0$, these terms are all related to the Einstein-Hilbert action, $S_{\text{ct},k}\sim \int (C_{0,k} +C_{1,k} \mathcal{R})$ with the Ricci scalar $\mathcal{R}$. This is best understood in terms of a spatial momentum cutoff $R_k(\vec p^{\,2})$ that decays at large momenta. Then, the flows are finite and resemble standard Euclidean flows, and the above renormalisation conditions emerge naturally for $R_k(\vec p^{\,2})\to Z_h\,k^2$.  

With the regularisation conditions in \cref{eq:RG-conditions}, the contributions from the single-graviton delta-peak read
\begin{align}
	\partial_t (\Gamma^{(hh)}_\text{TT}+S^{(hh)}_{\text{ct},\text{TT},k})\Big|_\text{tadpole}  =&\,0 \,,\notag \\[1ex]
	\partial_t (\Gamma^{(hh)}_\text{TT}+S^{(hh)}_{\text{ct},\text{TT},k})\Big|_\text{3-point}  (\tilde p = p/m_h) =&\,\frac{g\,m_h^2 (2-\eta_h)}{18 \pi ^2} \left(-84 + 26 \tilde p ^2+\frac{3 \left(11 \tilde p ^4-8 \tilde p ^2+56\right) \arccosh(1+\tilde p^2/2)}{\tilde p  \sqrt{\tilde p ^2+4}}\right) \,, \notag\\[2ex]
	\partial_t (\Gamma^{(hh)}_\text{TT}+S^{(hh)}_{\text{ct},\text{TT},k})\Big|_\text{ghost}   (\hat p = p/k)  =&\,\frac{2\,g\, k^2}{3 \pi} \left(30+ 7 \hat p^2-\frac{3 \left(\hat p^4+8 \hat p^2+20\right) \arccosh(1+\hat p^2/2)}{\hat p \sqrt{\hat p^2+4}}\right)\,.
\end{align}
Note that the graviton diagram only depends on $\tilde p = p/m_h$ with $m_h^2 = k ^2(1+\mu)$, while the  ghost diagram only depends on $\hat p = p/k$. The tadpole contribution is vanishing as expected from a massless tadpole diagram in dimensional regularisation. The structure of the graviton and ghost solution is identical with a characteristic $\arccosh$ contribution.

From the above equations, we can extract the contributions to the anomalous dimension, $\eta_h = \eta_h|_\text{3-point} + \eta_h|_\text{ghost}$,
\begin{align}
	\label{eq:flow-etah-explicit}
	\eta_h\Big|_\text{3-point} =&\, - g (2 -\eta_h) \frac{5 \pi  \sqrt{3}+147}{54\pi}\,, \notag\\[1ex]
	\eta_h\Big|_\text{ghost} =&\, -\frac{2 g}{3 \pi \hat m_h^2( 4 - \hat m_h^2)} \left( 60+ 4\hat  m_h^2-4 \hat m_h^4- \frac{3 \hat m_h^2 \left(\hat m_h^6-6 \hat m_h^4-4 \hat m_h^2+40\right) \arccosh\!\left(1-\hat m_h^2/2\right)}{ \sqrt{-\hat  m_h^4} \sqrt{\hat m_h^2(4-\hat m_h^2)}}	\right),
\end{align}
where $\hat m_h = m_h/k$, as well as to the graviton mass parameter, $\partial_t m_h^2 = \partial_t m_h^2|_\text{3-point} + \partial_t m_h^2|_\text{ghost} $,
\begin{align}
	\label{eq:flow-mh-explicit}
	\partial_t m_h^2\Big|_\text{3-point} =&\,g\, k^2 (2 -\eta_h) \frac{5 \left(5 \sqrt{3} \pi -22\right) \hat m_h^2}{18 \pi} \,,\notag \\[1ex]
	\partial_t m_h ^2\Big|_\text{ghost}=&\, \frac{2 g \,k^2 }{3 \pi } \left(30-7 \hat m_h^2 +\sqrt{\frac{\hat m_h^2}{4- \hat m_h^2}}\frac{3 \left(\hat m_h^4-8 \hat m_h^2+20\right) \arccosh\!\left(1-\hat m_h^2/2\right)}{\hat m_h^2 }\right).
\end{align}
Note that the ghost contributions in \cref{eq:flow-etah-explicit,eq:flow-mh-explicit} are only well defined for $\hat m_h^2 < 2$, which corresponds to $\mu< 3$. The flow of the multi-graviton continuum $\partial_t f_h = \partial_t f_{h,\text{3-point}} + \partial_t f_{h,\text{ghost}}$ is given by 
\begin{align}
	\label{eq:fh-flow}
	\partial_t f_{h,\text{3-point}}(\lambda)=&\, g (2 -\eta_h)\frac{56 m_h^4+8 m_h^2 \lambda^2+11 \lambda^4}{3 \lambda \left(m_h^2-\lambda^2\right)^2 \sqrt{\lambda^2-4 m_h^2}} \,\theta\!\left(\lambda^2-4 m_h^2\right)\,, \notag \\[2ex]
	\partial_t f_{h,\text{ghost}}(\lambda) =&\, -4 g \frac{20 k^4-8 k^2 \lambda^2+\lambda^4 }{\lambda \left(m_h^2-\lambda^2\right)^2 \sqrt{\lambda^2-4 k^2}} \,\theta\!\left(\lambda^2-4 k^2\right)\,.
\end{align}
The flow equation for the Newton coupling is taken from the Euclidean graviton three-point function at vanishing momentum \cite{Christiansen:2015rva, Denz:2016qks}. It reads in the given approximation
\begin{align}
	\label{eq:dt-G}
	\partial_t g = (2 + 3 \eta_h)\, g + \frac{g^2}{\pi}\! \left(-\frac{47 (6-\eta_h)}{114 (1+\mu)^2}+\frac{5 \left(8-\eta_h\right)}{38 (1+\mu)^3}+\frac{49 (10-\eta_h)}{570 (1+\mu)^4}-\frac{598}{285 (1+\mu)^5}-\frac{5}{19 }\right).
\end{align}
In contrast to the Lorentzian flow of the graviton two-point function, this flow is not on-shell, which can be seen from threshold terms such as $1/(1+\mu)^n$ in the flow. Furthermore, the flow was not obtained with a CS cutoff \cref{eq:CS-Cutoff}, but with a standard momentum cutoff $R_k(p)\propto (k^2 - p^2) \theta(k^2-p^2)$, \cite{Litim:2000ci}. The latter leads to factors like $(6-\eta_h)$ instead of $(2-\eta_h)$ typical for a CS cutoff. Despite these differences, the flows should be qualitatively compatible as long as $\mu$ is not close to the threshold $\mu=-1$ and $\eta_h$ remains small enough.

In the flow equations \cref{eq:flow-etah-explicit,eq:flow-mh-explicit,eq:fh-flow}, we have neglected the contribution of the multi-graviton continuum $f_h$ on the right-hand side since they are typically subleading. They read schematically,
\begin{align}
	\partial_t f_{h,\text{higher-order}}(\lambda) =&\,  \int_{2 m_h}^\infty \frac{\lambda_1 \mathrm d \lambda_1}{\pi} f_h(\lambda_1) F_1(\lambda,\lambda_1, m_h) +  \int_{2 m_h}^\infty \frac{\lambda_1  \mathrm d \lambda_1}{\pi} \frac{\lambda_2 \mathrm d \lambda_2}{\pi} f_h(\lambda_1) f_h(\lambda_2) F_2(\lambda,\lambda_1, \lambda_2, m_h) \notag \\[2ex]
	&\,+  \int_{2 m_h}^\infty \frac{\lambda_1  \mathrm d \lambda_1}{\pi} \frac{\lambda_2 \mathrm d \lambda_2}{\pi}\frac{\lambda_3 \mathrm d \lambda_3}{\pi} f_h(\lambda_1) f_h(\lambda_2)f_h(\lambda_3) F_3(\lambda,\lambda_1, \lambda_2,\lambda_3, m_h) \,,
\end{align}
and similarly for $\partial_t \mu$ and $\eta_h$.

\begin{figure}[t!]
	\includegraphics[width=.4\linewidth]{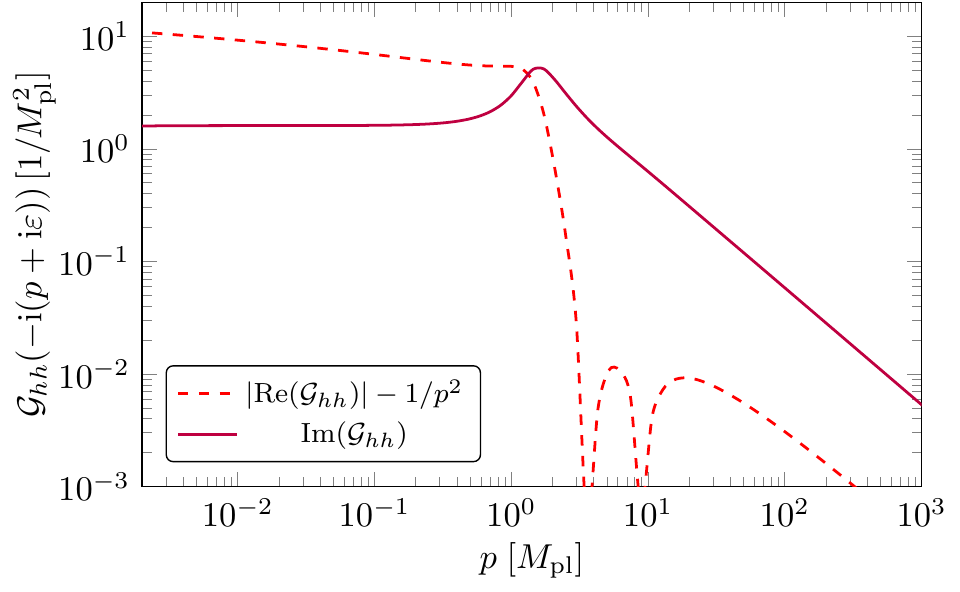}
	\caption{Real and imaginary part of the graviton propagator on the timelike axis.	
	}
	\label{fig:Lorentz-prop}
\end{figure}

\subsection{Propagator in the complex plane}\label{sec:complex-plane}

With the spectral function displayed in \cref{fig:rho_h-best}, we can compute the propagator in the whole complex plane, see \cref{eq:KS-L}. In our convention, fully real $p$ (or fully imaginary $\lambda$) are Euclidean, while fully imaginary $p$ (or fully real $\lambda$) are Lorentzian. As usual, we have a branch cut on the Lorentzian axis. The real and imaginary part of the propagator in the complex plane is displayed in \cref{fig:3d-prop}. The branch cut on the Lorentzian axis is clearly visible in the imaginary part of the propagator. Note also that the imaginary part on the Euclidean axis is exactly vanishing. The real and imaginary part of the propagator on the timelike axis is displayed in \cref{fig:Lorentz-prop}. The imaginary part is trivially related to the spectral function, see \cref{eq:rho-G}. The real part starts out positive for small momenta, becomes negative around the Planck scale, and then positive again around ten times the Planck scale.

\subsection{Analytic approximation}\label{sec:parameters}

\begin{figure}[t!]
	\includegraphics[width=.49\linewidth]{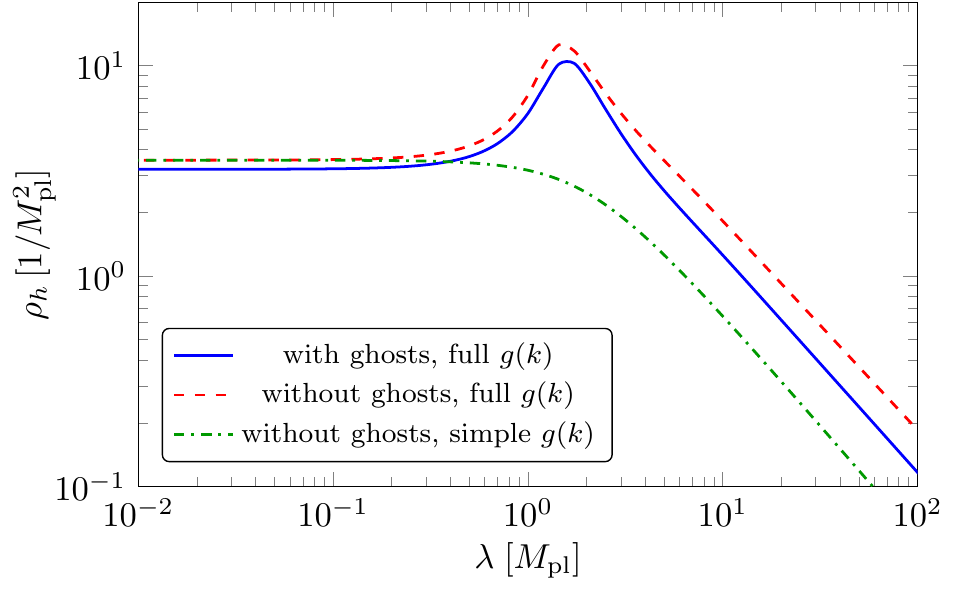}\hfill
	\includegraphics[width=.49\linewidth]{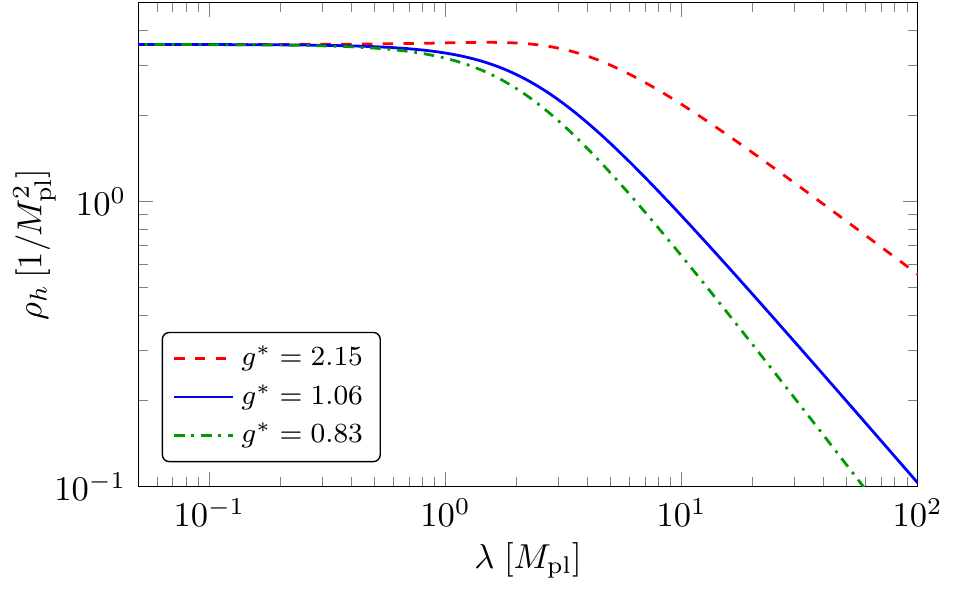}
	\caption{Spectral function of the fluctuation graviton in different approximations for $g^*=1.06$ (left) and in the analytical approximation for different fixed-point values of the Newton coupling (right). Here, $g^*=1.06$ is used in the present work, $g^*=2.15$ has been used for the reconstruction of the graviton spectral function in \cite{Bonanno:2021squ}, and $g^*=0.83$ is the fixed point value in the sophisticated Euclidean computation \cite{Denz:2016qks}. 
	}
	\label{fig:rho_h-approx}
\end{figure}

In this appendix, we summarise the flows and solutions in the analytic approximation, which we use for the computation of the spectral function at a finite cosmological constant. In this approximation, we neglect the ghost contributions and use a simplified trajectory for the Newton coupling, see \cref{eq:gk-simple}. The on-shell anomalous dimension reads 
\begin{align}
	\label{eq:flow-eta-app}
	\eta_{h}=\frac{2 g }{2 c_\eta  + g}\,,
	\qquad \text{with}\qquad 
	c_\eta =\frac{27\pi }{147 + 5\sqrt{3} \pi} \approx 0.49\,,
\end{align}
while  the flow of the on-shell graviton mass parameter is given by 
\begin{align}
	\label{eq:flow-mu-app}
	\partial_t \mu= -2 \mu - \eta_h+g(1+\mu)(2 - \eta_h) \frac{5 \left( 5\sqrt{3}\pi - 22 \right) }{18\pi} \,,
\end{align}
with the fixed point 
\begin{align}\label{eq:muh*}
	\mu^*=\frac{-g^* }{c_\mu  + g^*} \,,
	\qquad \text{with}\qquad 
	c_\mu =\frac{54 \pi }{477 - 70 \sqrt{3} \pi} \approx 1.77\,,
\end{align}
c.f.~\cref{eq:FP-mu-simple}. The flow equations \cref{eq:flow-eta-app,eq:flow-mu-app} have analytic solutions, c.f.~\cref{eq:mu-sol}, 
\begin{align}
	\label{eq:mu-sol-app}
	Z_{h}(k) = \left( 1 + \frac{1}{c_\eta\,\eta_h^*} \frac{k^2}{M_\text{pl}^2} \right)^{\!\!-\frac12 {\eta_h^*}}\,, \qquad \qquad 
	\mu(k)= \mu^* -\frac{2 \Lambda}{k^2} + \frac{c_1 M_\text{pl}^2-2\Lambda}{k^2} \left[Z_h(k)^{- c_2}-1\right],
\end{align}
with 
\begin{align}
	c_1 =  \frac{\frac{18 \pi }{5 \left(5 \sqrt{3} \pi -22\right)}  g^*}{\frac{54 \pi  \left(5 \sqrt{3} \pi -22\right)}{3925 \pi  \sqrt{3}-10494-1050 \pi ^2}+g^*} \approx \frac{2.17\, g^* }{1.77  + g^*} \,, 
	\qquad 	\qquad 
	c_2 = \frac{15 \left(5 \sqrt{3} \pi -22\right)}{5 \pi  \sqrt{3}+147} \approx 0.45 \,.
\end{align}
The flow of the multi-graviton spectrum can be integrated numerically on the analytic trajectories \cref{eq:mu-sol-app}. This allows us to understand the dependence of the spectral function on the IR cosmological constant, see \cref{fig:finite-Lambda} in the main text, as well as on the fixed-point value of the Newton coupling, see the right panel of \cref{fig:rho_h-approx}. In \cref{fig:rho_h-approx}, we used $g^*=1.06$ as in the main text as well as $g^*=2.15$, which was the fixed point value in \cite{Bonanno:2021squ}, and $g^*=0.83$, the fixed-point value from \cite{Denz:2016qks}. The fixed-point value of the Newton coupling changes the UV slope of the spectral function since the slope is proportional to $\sim \lambda^{\eta_h^*-2}$. The IR behaviour is untouched since it is related to the universal IR logarithmic branch cuts of the propagator. For $g^*=1.06$, we provide a simple analytic fit to the spectral function, which reads
	\begin{align}
		\rho_h(\lambda) = 2 \pi \delta(\lambda) + \theta(\lambda) \left( \frac{3.56}{0.92 \lambda ^2+0.82 \lambda +1}+\frac{8.50 \lambda ^{1.04}}{\lambda ^2+0.66 \lambda +2.54}\right).
	\end{align}
In the left panel of \cref{fig:rho_h-approx}, we compare the analytic approximation to the full solution. We can see that the negligence of the ghost contributions has only a small quantitative effect. The difference of the simplified trajectory for the Newton coupling \cref{eq:gk-simple} compared to the trajectory from the flow of the graviton three-point function \cref{eq:dt-G} is clearly visible around the Planck scale. While the simplified trajectory has no features at that scale, the full solution features a spike which can be traced back to the complex conjugated nature of the critical exponents.

\end{document}